\documentclass[twocolumn,aps,prl,showpacs,preprintnumbers,amsmath,amssymb]{revtex4}
\usepackage{amssymb}
\usepackage{graphics}
\usepackage{epsfig}
\usepackage{graphicx}    
\usepackage{dcolumn}     
\usepackage{amsthm}
\usepackage{bm}          
\usepackage{amsbsy}
\usepackage{color}
\makeatletter
    
    \newcommand{\Rmnum}[1]{\expandafter\@slowromancap\romannumeral #1@}
  \makeatother
\begin{document}
\title{All-magnonic spin-transfer torque and domain wall propagation}
\author{P. Yan$^{1}$, X. S. Wang$^{1}$,}
\author{X. R. Wang$^{1,2}$}
\email[Corresponding author: ]{phxwan@ust.hk}
\affiliation{$^1$Physics Department, The Hong Kong University of
Science and Technology,
Clear Water Bay, Hong Kong SAR, China \\
$^2$School of Physics, Wuhan University, Wuhan, China}

\begin{abstract}
The spin-wave transportation through a transverse magnetic domain
wall (DW) in a magnetic nanowire is studied. It is found that the
spin wave passes through a DW without reflection. A magnon, the
quantum of the spin wave, carries opposite spins on the two sides of
the DW. As a result, there is a spin angular momentum transfer from
the propagating magnons to the DW. This magnonic spin-transfer
torque can efficiently drive a DW to propagate in the opposite
direction to that of the spin wave.
\end{abstract}
 \maketitle
Magnetic domain wall (DW) propagation along nanowires has attracted
much attention in recent years \cite{Cowburn,Parkin,Yamaguchi,
Hayashi,Yanpeng,Wang,Han,Jamali} because of its fundamental interest
and potential applications \cite{Cowburn,Parkin}. So far, a
spin-polarized electric current and/or magnetic field including a
microwave \cite{Yanpeng} are the two known control parameters for
manipulating DW propagation along nanowires: A magnetic DW
propagates along a wire under a static magnetic field because of
energy dissipation \cite{Wang} while a DW moves under an electric
current because of spin-transfer torque (STT) \cite{Slon1,Berger}.
In terms of spintronic applications based on an electron spin
current STT, the Joule heating due to the excessive high critical
current density \cite{Yamaguchi,Hayashi} is a bottleneck. Thus, it
should be very interesting and important both academically and
technologically if one can find other effective control methods and
principles for DW manipulation in magnetic nanowires.

Both electrons and magnons, quanta of spin waves, carry spins. A
magnon is a spin-1 object with an angular momentum of $\hbar$
\cite{Stohr}. Similar to the STT from electrons to magnetization, a
STT from magnons to magnetization should in principle exist. Indeed,
polarized electric current generated by heat induced magnons in spin
valves was predicted theoretically \cite{Gerrit1,Slon2} and was
confirmed experimentally \cite{Yu}. The interaction between spin
waves and DW had also been investigated quantum mechanically
\cite{Yuan} and classically \cite{Hertel,Bayer,Macke}. The
time-dependent Schr\"{o}dinger equation was used \cite{Yuan} to show
that a DW is stable when it interacts with a spin wave, and the spin
wave is reflected by the DW which is different from our finding
below. Of course, they did not study magnonic STT. However, the
phase change of spin waves after passing through a DW was predicted.
In terms of STT directly from magnons, the question is how one can
facilitate a spin exchange between magnons and magnetization. In
this Letter, we show that the spin wave inside a DW satisfies a
Schr\"{o}dinger equation with a reflectionless potential well. A
magnon changes its spin by $2\hbar$ (the magnon spin flips from
$-\hbar$ to $\hbar$) after passing through the DW, as shown in Fig.
1. This angular momentum is absorbed by the DW, resulting in
propagation of the DW in the opposite direction to that of spin-wave
propagation. The validity of these findings is verified by solving
the Landau-Lifshitz-Gilbert (LLG) equation numerically in a
one-dimensional nanowire with material parameters of ferrimagnet
yttrium iron garnet (YIG). The frequency and the field dependences
of the DW propagation speed are also studied.

Consider a head-to-head DW in a magnetic nanowire whose easy axis
defined as the $z$ axis is along the wire as shown in Fig. 1, the
magnetization dynamics is described by the LLG equation \cite{Wang},
\begin{equation}
\frac{\partial \mathbf{m}}{\partial t}=-\mathbf{m}\times \mathbf{h}_{\text{%
eff}}+\alpha \mathbf{m}\times \frac{\partial \mathbf{m}}{\partial t},
\label{LLG}
\end{equation}%
where $\mathbf{m}$ is the unit direction of local magnetization
$\mathbf{M}=\mathbf{m}M_s$ with a saturation magnetization $M_s$,
$\alpha$ is the phenomenological Gilbert damping constant, and
$\mathbf{h}_{\text{eff}}$ is the effective magnetic field consisting
of anisotropy and exchange fields in the unit of $M_s$. $t$ is
normalized by $\left( \gamma M_s\right) ^{-1}$ and $\gamma$ is the
gyromagnetic ratio. For simplicity, we consider a uniaxial wire with
$\mathbf{h}_{\text{eff}}=Km_z\hat{z}+A\frac
{\partial^2\mathbf{m}}{\partial z^2}$ where $m_z$ is the $z$
component of $\mathbf{m},$ and $K$ and $A$ are the anisotropy and
exchange coefficients, respectively. In the spherical coordinates of
polar angle $\theta$ and azimuthal angle $\phi$, $\mathbf{m}=\left(
\sin\theta \cos\phi ,\sin\theta \sin\phi ,\cos\theta \right) $. For
a static DW, $\mathbf{m}= \mathbf{m}_0$ is given by the Walker
profile $\tan\frac{\theta_0} {2} =\exp\left(
\frac{z}{\Delta}\right)$ and lies in a fixed plane, say the $y$-$z$
plane ($\phi_0=\pi /2$), where $\Delta =\sqrt{A/K}$ is the DW width
\cite{Walker}.

\begin{figure}[tbph]
\begin{center}
\includegraphics[width=8.5cm]{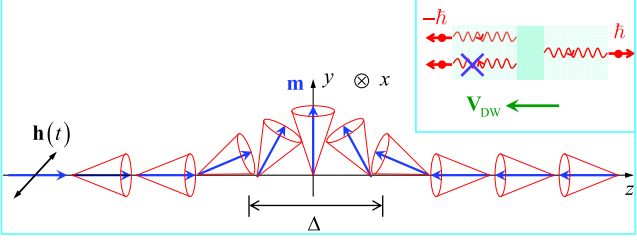}
\end{center}
\caption{(Color online) Illustration of a transverse DW structure
whose $\mathbf{m}$ is denoted by the (blue) arrows. The spin wave
(magnon) is a small amplitude precession of $\mathbf{m}$
(represented by the red cones) around the static DW. A linearly
polarized microwave $\mathbf{h}\left( t\right) $ is applied in a
small region on the left side of the DW so that the generated spin
wave propagates through the DW from the left. $\Delta$ is the DW
width. Inset: The magnons (wavy lines with arrows indicating the
propagating directions) pass through the DW (represented by the
rectangular block) from the left to the right without any
reflection. The magnon spin (solid circle with an arrow) is $-\hbar$
on the left side of the DW and $\hbar $ on the right side. The
magnonic STT drives the DW propagating to the opposite direction
(green arrow) of the spin wave, with the velocity
$\mathbf{V_{\text{DW}}}$.}
\end{figure}

To derive the equation of motion for the spin wave, a small
fluctuation of $\mathbf{m}$ around $\mathbf{m}_0$ is expressed in
terms of unit directions $\hat{e}_r,$ $\hat{e}_\theta$, and
$\hat{e}_\phi$ defined by $\mathbf{m}_{0}$,
\begin{equation}
\mathbf{m}\doteq \hat{e}_{r}+\left[ m_{\theta }\left( z\right) \hat{e}%
_{\theta }+m_{\phi }\left( z\right) \hat{e}_{\phi }\right] e^{-i\omega t},
\label{Fluctuation}
\end{equation}
where $\omega$ is the spin-wave frequency. $m_\theta$ and $m_\phi$
are small, $\sqrt{m_\theta^2+m_\phi^2}\ll 1$. Substituting Eq.
\eqref{Fluctuation} into Eq. \eqref{LLG} and neglecting the
higher-order terms, such as $m_\theta^2,$ $m_\theta m_\theta^ \prime
,$ $m_\theta m_\phi ,$ etc., ($^\prime$ denotes the derivative in
$z$), we obtain, in the absence of the damping,
\begin{eqnarray}
-i\omega m_\theta &=&Am_\phi^{\prime\prime}+K\left( 2\sin^2\theta
_{0}-1\right) m_{\phi },  \label{mt} \\
i\omega m_\phi &=&Am_\theta^{\prime\prime}+K\left( 2\sin^2\theta
_{0}-1\right) m_{\theta }.  \label{mp}
\end{eqnarray}%
Defining $\varphi =m_\theta -im_\phi$, \eqref{mt} and \eqref{mp} can
be recasted as
\begin{equation}
q^2\varphi \left( \xi\right) =\left[ -\frac{d^2}{d\xi^2}-2\text{sech}
^{2}\xi \right] \varphi \left( \xi \right) ,  \label{Schrodinger}
\end{equation}%
with $\xi =\frac{z}{\Delta}$, and $q^2=\frac{\omega}{K}-1$. This is
a Schr\"{o}dinger equation with propagating waves
\cite{Thiele,Dodd},
\begin{equation}
\varphi \left( \xi \right) =\rho \frac{\text{tanh}\xi -iq}{-iq-1}e^{iq\xi },
\label{wavefunction}
\end{equation}%
where $\rho$ is the spin-wave amplitude. Equation
\eqref{Schrodinger} also supports a bound state of $\varphi\left(
\xi\right) =\frac{1}{2}$sech$\xi$ for $q=-i$ ($\omega =0$)
\cite{Thiele,Dodd}.
Equation \eqref{wavefunction} describes propagating spin waves
without reflection, and takes an asymptotic form of $\varphi\left(
\xi\rightarrow -\infty \right) =\rho e^{iq\xi}$ and $\varphi\left(
\xi\rightarrow +\infty \right) =-\rho\frac{1-iq}{1+iq} e^{iq\xi }.$
The spin wave maintains its amplitude and only captures an extra
phase after passing through the DW. Interestingly, this phase shift
is indeed observed in recent calculations \cite{Hertel,Bayer,Macke}.
The above result is very robust, and holds even with the extra
Dzyaloshinskii-Moriya interaction \cite{DM,Oleg}
$D\mathbf{m}\cdot\left( \hat{z}\times\frac
{\partial\mathbf{m}}{\partial z}\right)$ in Eq. \eqref{LLG}.

A very interesting consequence of the above results is schematically
illustrated in the inset of Fig. 1: The magnons whose spins point to
the left (opposite to the magnetization of the left domain) are
injected into the DW from the left. The magnons transmit completely
through the DW with their spins reversed (to the right). The change
of magnon spins should be transferred to the DW, an all-magnonic
STT. Thus the DW propagates to the left, opposite to the magnon
propagation. One can also understand this result directly from Eq.
\eqref{LLG}. In the absence of damping, Eq. \eqref{LLG} can be cast
as
\begin{equation}
\frac{\partial \mathbf{m}}{\partial t}=-\mathbf{m}\times Km_z\hat{z}
-\frac{\partial }{\partial z}\mathbf{J},  \label{continuity}
\end{equation}%
where
$\mathbf{J}=A\mathbf{m}\times\frac{\partial\mathbf{m}}{\partial z}$
is the magnetization current, also called spin-wave spin current
\cite{Saitoh}. The $z$ component of Eq. \eqref{continuity} is
conserved so that $\partial _{t}m_{z}+\partial _{z}J_{z}=0$, where
$J_z$ is the $z$ component of $\mathbf{J}.$ In terms of $\varphi $,
$J_{z}=\frac{A}{2i\Delta }\left( \varphi \partial _{\xi }\varphi
^{\ast }-\varphi ^{\ast }\partial _{\xi }\varphi \right) \cos \theta
_{0}$ in the two domains. For the propagating spin wave
\eqref{wavefunction}, $J_z=-A\rho^2k$ in the far left of the wire
($z\rightarrow -\infty$ and $\theta_0=0$), while $J_z=A\rho^2k$ in
the far right ($z\rightarrow\infty$ and $\theta _0=\pi$), where
$k=q/\Delta$ is the spin-wave vector in real space. The spin current
changes its sign after passing through the DW, and results in an
all-magnonic STT on the DW. Thus, in order to absorb this torque,
the DW must propagate to the left with the velocity
$\mathbf{V_{\text{DW}}}=-\frac{\rho^2} {2} V_{g}\hat{z}$, where
$V_{g}=\partial \omega /\partial k=2Ak$ is the group velocity.

To test the validity of these findings in the realistic situation
when both damping and transverse anisotropy are present, we solve
Eq. \eqref{LLG} numerically in a one-dimensional magnetic nanowire.
In the simulations, the time, length, and field amplitude are in the
units of $\left( \gamma M_s\right)^{-1},$ $\sqrt{A/M_s},$ and $M_s,$
respectively, so that velocity is in the unit of
$\gamma\sqrt{AM_s}$. If one uses the YIG parameters: $M_s=0.194
\times10^6$ A/m, $K=0.388 \times10^5$ A/m, and $A=0.328\times
10^{-10}$ A m \cite{Krawczyk}, these units are $1.46\times10^ {-10}$
s, $13$ nm, $2.44\times10^{3}$ Oe, and $89$ m/s. The wire length is
chosen to be $1000$ (from $z=-500$ to $z=500$) with open boundary
conditions and a transverse DW is initially placed at the center of
the wire. Spin waves are generated by applying an external
sinusoidal magnetic field
$\mathbf{h}\left(t\right)=h_0\sin\left(\Omega t\right)\hat{x}$ of
frequency $\Omega$ and amplitude $h_0$ locally in the region of
[$-60$,$-55$] in the left side of the wire. Thus, the spin wave (may
not be monochromatic as explained later) propagates from the left to
the right as illustrated in Fig. 1. We solve Eq. \eqref{LLG}
numerically by using the standard method of lines. The space is
divided into small meshes of size $0.05$ and an adaptive time-step
control is used for the time evolution of the magnetization. In
terms of YIG parameters, the geometry of our nanowire is 0.65 nm
$\times$ 0.65 nm in cross section and 13 $\mu$m in length. The DW
will move under the influence of the spin wave. The spatial-temporal
dependence of $m_z$ is used to locate the DW center which, in turn,
is used to extract the DW velocity.
\begin{figure}[tbph]
\begin{center}
\includegraphics[width=8.5cm]{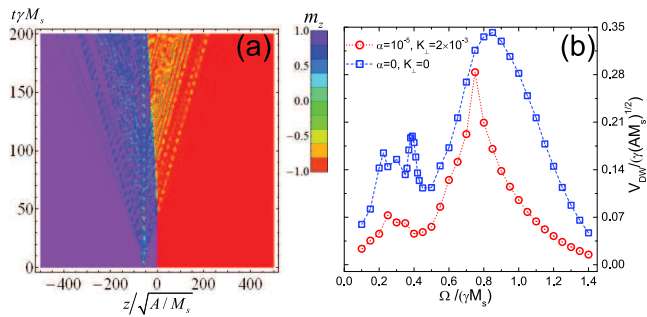}
\end{center}
\caption{(Color) (a) Density plot of $m_{z}$ in the $z$-$t$ plane at
frequency $\Omega =0.75$ for $\alpha=10^{-5}$ and $K_{\perp}=2\times
10^{-3}$ (YIG parameters). The center of the DW is initially at
$z=0$. The spin wave is generated in the region of $z\in [-60,-55]$.
(b) The frequency dependence of DW velocity. The red circles are for
YIG parameters, and the blue squares are the results of the case
without damping and transverse magnetic anisotropy.}
\end{figure}

Below, we present our simulations for a set of realistic material
parameters of YIG: $\alpha=10^{-5}$ and $K_{\perp}=2\times 10^{-3}$.
We present also the simulation results when both damping and
transverse magnetic anisotropy are absent in order to show the
quantitative effects of damping and transverse anisotropy although
the qualitative results are the same. Figure 2(a) is the numerical
results of the spatial-temporal dependence of $m_z$ at $h_0=1$ and
$\Omega =0.75$ (optimal frequency explained later) for YIG. The
simulations show the following interesting results. First, spin
waves are generated by the external field. The spin waves propagate
to both sides of the wire, resulting in the parallel strap pattern
in the density plot of $m_z$. Second, when the spin waves reach the
DW at $z=0$, the DW starts to move towards the left, opposite to the
spin-wave propagating direction.
The straight trajectory of the DW center before hitting the wave
source indicates that the DW propagation speed is almost a constant.
Third, the slopes of the spin-wave straps and DW trajectory tell us
that the DW propagation speed is smaller than the spin-wave group
velocity, a reasonable result that is consistent with our picture.
It is also clear that there is no reflection when the spin waves
pass through the DW in the presence of both damping and transverse
magnetic anisotropy. This is highly nontrivial since it is not so
clear from our earlier analysis. We will present further evidence
for this finding. Figure 2(b) shows the frequency dependence of DW
velocity at $h_0=1$ for both YIG parameters (circles) and the case
without damping and transverse magnetic anisotropy (squares). The
error bars are smaller than the symbol sizes. The complicated and
irregular frequency dependence of the DW velocity at low frequency
is probably related to the observation of the polychromatic
spin-wave generation. At a large enough frequency ($\Omega >0.55$),
the excited spin wave is almost monochromatic with the same
frequency as the oscillating field. These curves show that the DW
propagation velocity is very sensitive to the microwave frequency.
In fact, there exists an optimal frequency at which the DW velocity
is maximal for a given set of parameters. For the cases shown in the
figure, the optimal frequencies are $\Omega =0.75$ in the presence
of damping and transverse magnetic anisotropy and a higher optimal
frequency $\Omega =0.85$ without them.
\begin{figure}[tbph]
\begin{center}
\includegraphics[width=7cm]{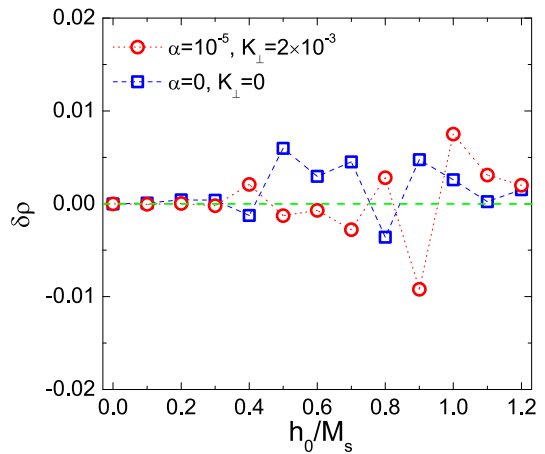}
\end{center}
\caption{(Color online) Field dependence of spin-wave amplitude
differences at two sites located in the opposite sides of the DW.
Red circles are for the YIG parameters at the optimal frequency
($\Omega =0.75$), and blue squares are the results without damping
and the transverse magnetic anisotropy also at its optimal frequency
($\Omega=0.85$).}
\end{figure}

The reflectionless property (total transmission) of the spin wave
through DW can also be verified through quantitative analysis of
spin-wave amplitude on the two opposite sides of the DW. If the spin
wave is monochromatic (a sinusoidal wave) and passes through the DW
without reflection, the difference of the spin-wave amplitudes on
the two sides of the DW is around zero. We evaluate the spin-wave
amplitude difference at $z=-160$ and $z=45$ at the same time,
denoted as $\delta\rho$. Figure 3 is the numerical results of
$\delta\rho$ as a function of microwave field $h_{0}$. Indeed,
$\delta\rho$ is almost zero (green dashed line) both with (circles)
and without (squares) damping and transverse magnetic anisotropy. Of
course, for nonmonochromatic (sum of many sinusoidal waves) spin
waves, the amplitudes on the two sides of the DW may be different at
any particular time due to the complicated interference of waves
with different frequencies. This is indeed the case for large $h_0$,
as shown by the oscillatory $\delta\rho$ around zero. The total
transmission of spin waves through a DW is an important property
because it results in a larger spin-wave spin current, and generates
a larger magnonic STT.
\begin{figure}[tbph]
\begin{center}
\includegraphics[width=8.5cm]{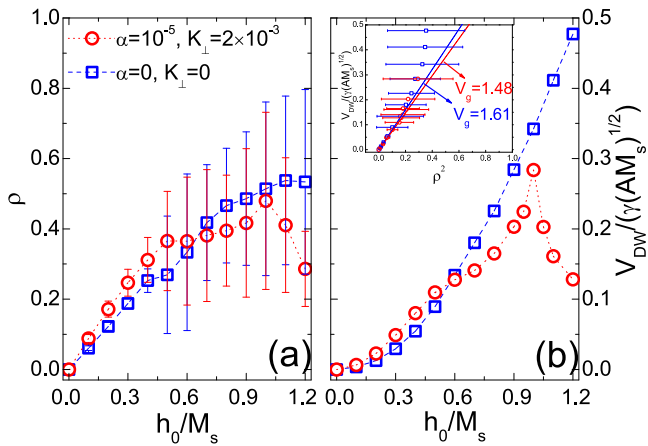}
\end{center}
\caption{(Color online) (a) Field dependence of spin-wave amplitude
for YIG parameters (red circles) at frequency $\Omega =0.75$ and the
case without damping and transverse magnetic anisotropy (blue
squares) at frequency $\Omega=0.85$. (b) Field dependence of the DW
velocity for the same cases as those in (a). Inset: DW velocity vs
the square of the spin-wave amplitude. Symbols are the simulation
data, and solid lines are $\frac{\rho^2} {2} V_{g}$ without any
fitting parameters.}
\end{figure}

Figure 4(a) shows the $h_0$ dependence of the spin-wave amplitude
$\rho$. It is almost linear at low fields both with (circles) and
without (squares) damping and transverse magnetic anisotropy. The
behavior is complicated at high fields, and a large error bar of
$\rho$ is observed, accompanying less regular and polychromatic
spin-wave generation. This also results in a large fluctuation of
$\rho$. The $h_{0}$ dependence of the DW velocity $V_{\text{DW}}$ is
shown in Fig. 4(b). It is nonmonotonic for the realistic situation
with YIG parameters (red circles) and almost quadratic for the case
without damping and transverse magnetic anisotropy (blue squares).
Although the field dependence of DW velocity is nonmonotonic, the
relationship between the DW velocity and the spin-wave amplitude is
much simpler. As shown in the inset of Fig. 4(b), the DW velocity
$V_{\text{DW}}$ is almost quadratic in $\rho$ both with (circles)
and without (squares) damping and transverse magnetic anisotropy. We
also plot $\frac{\rho^2} {2} V_{g}$ (solid lines) without any
fitting parameters, where $V_{g}=1.48$ at $\Omega =0.75$ for the YIG
case and $V_{g}=1.61$ at $\Omega=0.85$ in the absence of damping and
transverse anisotropy, and $\rho$ is calculated numerically.
Although there is no reason why the early velocity formula derived
under the approximation of zero damping for uniaxial wire and small
spin-wave amplitude should be applicable to the realistic case when
both damping and transverse magnetic anisotropy are presented, the
theoretical formula is, in fact, not too far from the numerical data
for both cases. Of course, it should not be surprising for the
deviation at large $\rho$ since quadratic $\rho$ dependence of
$V_{\text{DW}}$ is derived based on the conservation of the $z$
component of angular momentum that does not hold for the generic
cases. Also the main purpose of the current study is to demonstrate
the principles rather than the exact mathematical expression of DW
velocity which can be the subject of future studies.

Most studies \cite{Gerrit1,Slon2,Yu,Saitoh,Gerrit2,Uchida} of
magnonic effects in nanomagnetism so far are about the conversion of
magnon spins with electron spins. Very often, it goes through the
Seebeck effect that involves both thermal and electronic transport.
Thus, like usual electronic STT, devices based on these effects must
also contain metallic parts so that Joule heating shall be present.
In contrast, the magnonic STT presented here does not require
electron transport. Devices based on this all-magnonic STT could be
made of magnetic insulators like YIG so that the Joule heating is,
in principle, avoided. It is also known that the stray field is
important in DW dynamics. Our results here are consistent with the
OOMMF \cite{OOMMF} simulations including this field. Remarkably,
these results are consistent with a phenomenological theory on
thermomagnonic STT proposed by Kovalev and Tserkovnyak
\cite{yaroslav}.

In conclusion, we proposed an all-magnonic spin-transfer torque
mechanism for magnetic domain wall manipulation in nanowires. This
spin-transfer torque can effectively drive a DW to propagate along
the wires. The propagation speed is sensitive to both microwave
frequency and its amplitude. There is an optimal frequency, order of
the usual ferromagnetic resonance frequency, at which DW propagating
speed is the fastest. All-magnonic STT should have advantages over
its electronic counterpart on energy consumption as well as on the
spin-transfer efficiency. It also opens the door for using magnetic
insulators in spintronic devices.





This work is supported by Hong Kong RGC Grants (No. 604109, No.
RPC11SC05, and No. HKUST17/CRF/08).


\begin{thebibliography}{99}
\bibitem{Cowburn} D.A. Allwood, G. Xiong, C.C. Faulkner, D. Atkinson, D.
Petit, and R.P. Cowburn, Science \textbf{309}, 1688 (2005).
\bibitem{Parkin} S.S.P. Parkin, M. Hayashi, and L. Thomas, Science \textbf{320}, 190 (2008).
\bibitem{Yanpeng} P. Yan and X.R. Wang, Phys. Rev. B \textbf{80}, 214426
(2009).
\bibitem{Wang} X.R. Wang, P. Yan, J. Lu, and C. He, Ann. Phys. (N. Y.)
\textbf{324}, 1815 (2009); X.R. Wang, P. Yan, and J. Lu, Europhys. Lett.
\textbf{86}, 67001 (2009).
\bibitem{Yamaguchi} A. Yamaguchi, T. Ono, S. Nasu, K. Miyake, K. Mibu, and T. Shinjo, Phys. Rev. Lett. \textbf{92}, 077205 (2004).
\bibitem{Hayashi} M. Hayashi, L. Thomas, Y.B. Bazaliy, C. Rettner, R.
Moriya, X. Jiang, and S.S.P. Parkin, Phys. Rev. Lett. \textbf{96},
197207 (2006).

\bibitem{Han} D.S. Han, S.K. Kim, J.Y. Lee, S.J. Hermsdoerfer, H.
Schultheiss, B. Leven, and B. Hillebrands, Appl. Phys. Lett.
\textbf{94}, 112502 (2009).

\bibitem{Jamali} M. Jamali, H. Yang, and K.J. Lee, Appl. Phys. Lett.
\textbf{96}, 242501 (2010).
\bibitem{Slon1} J. Slonczewski, J. Magn. Magn. Mater. \textbf{159}, L1
(1996).
\bibitem{Berger} L. Berger, Phys. Rev. B \textbf{54}, 9353 (1996).
\bibitem{Stohr} J. St\"{o}hr and H.C. Siegmann, \emph{Magnetism: From
Fundamentals to Nanoscale Dynamics} (Springer-Verlag, Berlin, 2006).

\bibitem{Gerrit1} M. Hatami, G.E.W. Bauer, Q. Zhang, and P.J. Kelly, Phys.
Rev. Lett. \textbf{99}, 066603 (2007).
\bibitem{Slon2} J.C. Slonczewski, Phys. Rev. B \textbf{82}, 054403 (2010).
\bibitem{Yu} H. Yu, S. Granville, D.P. Yu, and J.P. Ansermet, Phys. Rev.
Lett. \textbf{104}, 146601 (2010).
\bibitem{Yuan} S. Yuan, H.D. Raedt, and S.
Miyashita, J. Phys. Soc. Jpn. \textbf{75}, 084703 (2006).
\bibitem{Hertel} R. Hertel, W. Wulfhekel, and J. Kirschner,
Phys. Rev. Lett. \textbf{93}, 257202 (2004).
\bibitem{Bayer} C. Bayer, H. Schultheiss, B.
Hillebrands, and R.L. Stamps, IEEE Trans. Magn. \textbf{41}, 3094
(2005).
\bibitem{Macke} S. Macke and D. Goll, J. Phys. Conf. Ser. \textbf{200}, 042015 (2010).
\bibitem{Walker} N.L. Schryer and L.R. Walker, J. Appl. Phys. \textbf{45},
5406 (1974).
\bibitem{Thiele} A.A. Thiele, Phys. Rev. B
\textbf{7}, 391 (1973).

\bibitem{Dodd} R.K. Dodd, \emph{Solitons and Nonlinear Wave equations}
(Academic Press, London, 1982).
\bibitem{DM} I. Dzyaloshinsky, J. Phys. Chem. Solids \textbf{4}, 241 (1958); T. Moriya, Phys. Rev. \textbf{120}, 91 (1960).

\bibitem{Oleg} O.A. Tretiakov and A. Abanov, Phys. Rev. Lett. \textbf{105}, 157201 (2010).

\bibitem{Saitoh} Y. Kajiwara, K. Harii, S. Takahashi, J. Ohe, K. Uchida, M. Mizuguchi, H. Umezawa, H. Kawai, K. Ando, K. Takanashi, S. Maekawa,
and E. Saitoh, Nature (London) \textbf{464}, 262 (2010).
\bibitem{Krawczyk} M. Krawczyk and H. Puszkarski, Cryst. Res.
Technol. \textbf{41}, 547 (2006).
\bibitem{Gerrit2} G.E.W. Bauer and Y. Tserkovnyak, Physics \textbf{4}, 40
(2011).
\bibitem{Uchida} K. Uchida, J. Xiao, H. Adachi, J. Ohe, S. Takahashi, J.
Ieda, T. Ota, Y. Kajiwara, H. Umezawa, H. Kawai, G.E.W. Bauer, S.
Maekawa, and E. Saitoh, Nature Mater. \textbf{9}, 894 (2010).
\bibitem{OOMMF} M.J. Donahue and D.G. Porter, National Institute of Standards and
Technology Interagency Report No. NISTIR 6376, 1999.
\bibitem{yaroslav} A.A. Kovalev and Y. Tserkovnyak, arXiv:1106.3135.
\end{thebibliography}
\end{document}